\documentstyle{article}
\def\1ad{\mbox{\normalsize $^1$}}
\def\2ad{\mbox{\normalsize $^2$}}
\def\3ad{\mbox{\normalsize $^3$}}
\def\4ad{\mbox{\normalsize $^4$}}
\def\5ad{\mbox{\normalsize $^5$}}
\def\6ad{\mbox{\normalsize $^6$}}
\def\7ad{\mbox{\normalsize $^7$}}
\def\8ad{\mbox{\normalsize $^8$}}
\def\makefront{\vspace*{1cm}\begin{center}
\def\newtitleline{\\ \vskip 5pt}
{\Large\bf\titleline}\\
\vskip 1truecm
{\large\bf\authors}\\
\vskip 5truemm
\addresses
\end{center}
\vskip 1truecm {\bf Abstract:} \abstracttext \vskip 1truecm}
\def\cN{{\cal N}}

\def\cM{{\cal M}}
\def\cK{{\cal K}}
\def\cQ{{\cal Q}}
\def\cG{{\cal G}}
\def\cU{{\mathcal U}}
\def\half{{1 \over 2}}
\def\IR{\relax{\rm I\kern-.18em R}}
\def\IC{\relax\,{\rm I\kern-.5em{\rm C}}}
\def\l{\lambda}

\def\s{\sigma}
\def\a{\alpha}
\def\b{\beta}
\def\L{\Lambda}
\def\S{\Sigma}

\def\beq{\begin{equation}}
\def\eeq{\end{equation}}
\begin{document}
%HU-EP-00/57
\def\titleline{
$N=2 \to N=1$ supergravity reduction in four dimensions
%\newtitleline
}
\def\authors{
Laura Andrianopoli \1ad }
\def\addresses{
\1ad Dip. Fisica, Politecnico di Torino,
 Corso Duca degli Abruzzi 24, I-10129 Torino \\
{\tt e-mail: andrianopoli@athena.polito.it} }
\def\abstracttext{We discuss the reduction of $N=2$ supergravity to $N=1$,
by a consistent truncation of the second gravitino multiplet.}
\large \makefront
%%%%%%%%%%%%%%%%%%%%%%%%%%%%%%%%%
%%%%%%%%%%%%%%%%%%%%%%%%%%%%%%%%%

It is well known that for globally supersymmetric theories with
particle content of spin $0,\half, 1$ any theory with $N$
supersymmetries can be regarded as a particular case of a theory
with a number $N'< N$ of supersymmetries \cite{st}. To prove this
it is sufficient to decompose the $N$ supersymmetry--extended
multiplets into $N'$-multiplets. Of course $N$-extended
supersymmetry is more restrictive than $N'< N$ supersymmetry
implying that the former will only allow some restricted couplings
of the latter. As we are going to show, the same argument does not
apply to supergravity theories \cite{lungo}. Indeed, let us
consider a standard $N$-extended supergravity with $N$ gravitini
and a given number of matter multiplets: then the $N'$-extended
supergravity obtained by reduction from the mother theory will no
longer be standard because a certain number $N-N'$ of spin $\frac
{3} {2}$ multiplets appear in the decomposition. To obtain a
standard $N'$-extended supergravity one must truncate out at least
the $N-N'$ spin $\frac {3} {2}$ multiplets and all the non-linear
couplings that they generate in the supergravity action.
\par
Here I will report on the truncation of $N=2$ matter-coupled
supergravity theory to $N=1$, showing that the reduction does
indeed imply the truncation of part of the matter content, besides
of the second gravitino multiplet. The present discussion is based
on \cite{lungo}. A more detailed analysis, with proofs of
consistency, and including the truncation of $N=8$ supergravity
down to general $N$ theories can be found in \cite{lungo} to which
I refer also for the notations and conventions and for a complete
list of references.
\par
 The supersymmetry reduction $N=2\to N=1$ is obtained by truncating
the $N=1$ spin $3/2$ multiplet containing the second gravitino
$\psi_{\mu 2}$ and the graviphoton.
\par
Let us write down the supersymmetry transformation laws of the
fermions for the $N=2$ theory, up to 3-fermions terms
\cite{abcdffm}:
\begin{eqnarray}
\delta\,\psi _{A \mu} &=& {\hat{\nabla}}_{\mu}\,\epsilon _A\,
 + \left ( {\rm i} \, g \,S_{AB}\eta _{\mu \nu}+
\epsilon_{AB} T^-_{\mu \nu} \right) \gamma^{\nu}\epsilon^B
 \label{gravtrasf} \\
\delta \,\lambda^{{\mathcal I}A}&=&
 {\rm i}\,\nabla _ {\mu}\, z^{{\mathcal I}}
\gamma^{\mu} \epsilon^A +G^{-{\mathcal I}}_{\mu \nu} \gamma^{\mu
\nu} \epsilon _B \epsilon^{AB}\,+\, gW^{{\mathcal I}AB}\epsilon _B
\label{gaugintrasf}\\
 \delta\,\zeta _{\alpha}&=&{\rm i}\,
{\mathcal U}^{B \beta}_{u}\, \nabla _{\mu}\,q^u \,\gamma^{\mu}
\epsilon^A \epsilon _{AB}\,C_{\alpha  \beta} \,+\,g
N_{\alpha}^A\,\epsilon _A \label{iperintrasf}
\end{eqnarray}
where: $\hat{\nabla}_{\mu}\,\epsilon _A = {\mathcal D}_\mu
\epsilon_A
 + \hat{\omega}_{\mu |
A}^{\phantom{\mu |A}B} \epsilon_B +\hat{\cal Q}_\mu \epsilon_A $,
 with the $SU(2)$ and $U(1)$ 1-form ``gauged'' connections
respectively given by: $
\hat{\omega}_{A}^{\phantom{A}B}={\omega}_{A}^{\phantom{A}B} +
g_{({\bf\L})} \, A^{{\bf\L}} \,P^x_{{\bf\L}}\,
(\s^x)_{A}^{\phantom{A}B}$; $\hat{\cal Q} = {\cal Q} +
g_{({\bf\L})}\, A^{{\bf\L}} \,P^0_{{\bf\L}} $,
${\omega}_{A}^{\phantom{A}B}$ and ${\cal Q} = -\frac{\rm i}2
(\partial_{{\mathcal I}} \cK d z^{{\mathcal I}} -
\partial_{\bar {\mathcal I}} \cK d \bar z^{\bar {\mathcal
I}})$ being the
 $SU(2)$ and $U(1)$ connections of the ungauged
theory.
  $T^-_{\mu\nu}$ appearing in the supersymmetry
transformation law of the $N=2$ left-handed gravitini is the
``dressed'' graviphoton defined as:
\begin{eqnarray}\label{gravif}
T^-_{\mu\nu} &\equiv & 2{\rm i} {\rm Im} {\cal N}_{{\bf \L } {\bf
\S } } L^{{\bf \S } }\Bigl[  F_{\mu\nu}^{{\bf \L }  -} + \mbox{3
fermions terms}
 \Bigr]
\end{eqnarray}
while
\begin{eqnarray}\label{vectors}
G^{{\mathcal I}-}_{\mu\nu} &\equiv & - g^{{\mathcal
I}\bar{\mathcal J}} {\rm Im } {\cal N}_{ {\bf \L }{\bf \S }
}\,\bar f^{{\bf \S } }_{\bar{\mathcal J}}
  \Bigl[  F_{\mu\nu}^{{\bf \L }  -} +   \mbox{3
fermions terms}
 \Bigr]
\end{eqnarray}
are the ``dressed'' field strengths of the vectors inside the
vector multiplets (the ``minus'' apex means taking the self-dual
part.). Moreover the fermionic shifts $S_{AB}$, $ W^{{\mathcal
I}\,AB}$ and $ N^A_{\alpha}$ are given in terms of the
prepotentials and Killing vectors of the quaternionic geometry as
follows:
\begin{eqnarray}\label{trapsi2}
S_{AB}&=& {\rm i} \frac {1}{2} P_{AB\, {\bf\L} } \,
 L^{{\bf\L} }
 \equiv {\rm i} \frac {1}{2} P^x_{{\bf\L}} \sigma^x_{AB}L^{{\bf\L}} \\
\label{tralam}
 W^{{\mathcal I}\,AB}&=& {\rm {i}} P^{AB}_{
{\bf\L} }\,g^{{\mathcal I}\bar {\mathcal J}} f^{{\bf\L} }_{\bar
{\mathcal J}} + \epsilon^{AB} k^{\mathcal I}_{{\bf\L} }{\overline
L}^{{\bf\L} }
\\
N^A_{\alpha}&=& 2\,{\mathcal U}^A_{\alpha u} \,k^u_{{\bf\L} }
{\overline L}^{{\bf\L} } \label{traqua1}
\end{eqnarray}
Since we are going to compare the $N=2$ reduced theory with the
standard $N=1$ supergravity, I also quote the supersymmetry
transformation laws of fermions in the latter theory
\cite{bw2},\cite{cfgv}. We have, up to 3-fermions terms:
 \\
{\bf $N=1$ transformation laws}
\begin{eqnarray}
\label{trapsi1}\delta \psi_{\bullet \mu} &=& {\cal D}_{\mu}
\epsilon_{\bullet}+ \frac{\rm i}2 \hat Q_\mu\epsilon_{\bullet}
+{\rm {i}} L(z, \bar z) \gamma_{\mu} \varepsilon^{\bullet} \\
\label{trachi1} \delta \chi^i &=& {\rm {i}}
\left(\partial_{\mu}z^i + g_{(\L )} A^\L_\mu k^i_\L \right)
\gamma^{\mu} \varepsilon_{\bullet}  + N^{i}\varepsilon_{\bullet}\\
\label{tralamb1}\delta \lambda^{\L }_{\bullet} &=& {\mathcal
{F}}_{\mu \nu}^{(-) \L } \gamma^{\mu \nu} \varepsilon_{\bullet }
+{\rm {i}} D^{\L } \varepsilon_{\bullet}
\end{eqnarray} where
$\hat \cQ$ is defined in a way analogous to the $N=2$ definition
 and:
\begin{eqnarray}\label{n1def}
L(z,\bar z)&=& W(z)\,e^{\frac {1}{2} {{\mathcal {K}}_{(1)}(z, \bar z)}}\,,\quad \nabla_{\bar\imath}L =0 \\
\label{defn} N^i &=& 2\, g^{i\bar\jmath} \,\nabla_{\bar\jmath}\, \bar L \\
 \label{dlambda} D^{\L } &=& - 2 ({\rm {Im}} f_{\L
\S })^{-1} P_{\S }(z,\bar z)
\end{eqnarray}
and $W(z),{\mathcal {K}}_{(1)}(z, \bar z) ,P_{\S }(z,\bar z),
f_{\L\S}(z)$ are the superpotential,  K\"{a}hler potential,
Killing prepotential and  vector kinetic matrix respectively
\cite{cfgv}, \cite{bw2}, \cite{bagger}. Note that for the
gravitino and gaugino fields we have denoted by a lower (upper)
dot left-handed (right-handed) chirality. For the spinors of the
chiral multiplets $\chi$, instead, left-handed (right-handed)
chirality is encoded via an
 upper holomorphic (antiholomorphic) world index
 ($\chi^i, \chi^{\bar\imath}$).
 %%%%%%%%%%%%%%%%%%%%%%%%%%%%%%%%%%%%%%%%%%%
%%%%%%%%%%%%%%%%%%%%%%%%%%%%%%%%%%%%%%%%%%
%%%%%%%%%%%%%%%%%%%%%%%%%%%%%%%%%%%%%%%%%%%
\par
 To perform the truncation
we set $A$=1 and 2 successively, putting $\psi_{2\mu}
=\epsilon_2=0$, and for the gravitino we get, from equation
(\ref{gravtrasf}):
\begin{equation}
\label{reductio} \delta\,\psi _{1 \mu} = {\mathcal D}_\mu
\epsilon_1  -\hat{\cal Q}_\mu \epsilon_1 - \hat{\omega}_{\mu |
1}^{\phantom{\mu |A}1} \epsilon_1  \,
 +  {\rm i} \, g \,S_{11}\eta _{\mu \nu}
 \gamma^{\nu}\epsilon^1
\end{equation}
while, for consistency:
\begin{equation}
\delta\,\psi _{2 \mu}\equiv 0 =  - \hat{\omega}_{\mu |
2}^{\phantom{\mu |A}1} \epsilon_1 +
  \left ({\rm i} \, g \,S_{21}\eta _{\mu \nu}-T^-_{\mu \nu}\right)
 \gamma^{\nu}\epsilon^1
\end{equation}
\par
Comparing (\ref{trapsi1}) with (\ref{reductio}), we learn that we
must identify $\psi_{1\mu} \equiv  \psi_{\bullet \mu}\,; \quad
\epsilon_1 \equiv
  \epsilon_{\bullet }$.
Furthermore,  $g\,S_{11}$ must be identified with the
superpotential of the $N=1$ theory, that is to the covariantly
holomorphic section $L$ of the $N=1$ K\"{a}hler-Hodge manifold.
Therefore we have \cite{ps} $
  L(q,z,\bar z)=\frac{\rm i}{2} g_{(\bf\L )} P^x_{\bf\L}
(\s^x)_{11} L^{\bf\L} = \frac{\rm i}{2} g_{(\bf\L )} \left(
P^1_{\bf\L} - {\rm i} P^2_{\bf\L} \right) L^{\bf\L}$. As it has
been shown in \cite{lungo}, after consistent reduction of the
special-K\"ahler manifold $\cM^{SK}$ and of the quaternionic
$\s$-model $\cM^Q$, $L$ is in fact a covariantly holomorphic
function of
 the K\"ahler coordinates $w^s$ of the reduced manifold
${\cal M}^{KH} \subset {\cal M}^Q$ and of some subset $z^i \in
\cM_R$ of the scalars $z^{\mathcal I}$ of the $N=2$
special-K\"ahler manifold $\cM^{SK}$.
\par
 Furthermore, for a consistent truncation we must
set to zero all the following bosonic structures:
\begin{eqnarray}
&&T^-_{\mu \nu} = 0 \label{tmunu}\\
&&\hat{\omega}_{\mu | 2}^{\phantom{\mu |A}1} =0\label{o21}\\
&&S_{21} =0\label{s}
\end{eqnarray}
\par
As it has been proven in \cite{lungo}, equations (\ref{tmunu}) and
(\ref{o21}) are ``orthogonality'' conditions  on the scalar
sectors of $N=2$ supergravity which imply a reduction of both
special K\"ahler manifold ($\cM^{SK}(n_V)$) \cite{dlv},\cite{str},
\cite{cadf}, \cite{abcdffm}
  and quaternionic  manifold
 ($\cM^{Q}(n_H)$), where $n_V$ and $n_H$ are the number of
 vector multiplets  and hypermultiplets respectively, and
  imply therefore a truncation of part of the matter multiplets.
 We note that similar orthogonality conditions,
 leading to the same reduction of the matter content of the theory,
 can be found through a complementary analysis,
  by looking at the 3-fermions terms in the transformation laws,
  instead of at the bosonic ones \cite{corto}.
\par
 Equation  (\ref{tmunu}) is satisfied by imposing a
suitable constraint on the set of vectors and of scalar sections
which can be retained in the reduction. Indeed, if we decompose
the index $\bf\L$ labelling the vectors into two disjoint sets
${\bf\L} \Rightarrow ( \L , X ), \L =1,\cdots ,n_V'=n_V -n_C;X =
0,1,\cdots ,n_C $, we may satisfy the relation (\ref{tmunu}) as an
``orthogonality relation'' between the subset $ \L$ running on the
retained vectors and the subset $X$ running on the retained scalar
sections. That is we set:
\begin{eqnarray}
F^{X}_{\mu\nu} &=& 0 ;\\
{\rm Im} \cN_{\L{\bf\S}} L^{\bf\S} &=& T_\L=0 \label{nl}
\end{eqnarray}
We note that if we delete the electric field strengths $F^{-X}$ we
must also delete their magnetic counterpart $\cG^-_{X}= \bar
\cN_{XY} F^{-Y} + \bar \cN_{X   \S} F^{- \S}=0$ so that we must
also impose $  \cN_{X  \S} =0$. Then, the constraint (\ref{nl})
reduces to ${\rm Im} \cN_{ \L \S} L^{ \S} =0$ which implies
\begin{equation}
 L^{ \S} =0\label{scalred1}
\end{equation}
since the vector-kinetic matrix ${\rm Im} \cN_{ \L \S}$ has to be
invertible. Note that condition (\ref{scalred1}) implies a
reduction of the $N=2$ scalar manifold $\cM^{SK} \to \cM_R$, since
it says that some coordinate dependent sections on $\cM^{SK}$ have
to be zero in the reduced theory. \\ Let us now quote the
reduction of the gauginos transformation law. When $\epsilon_2 =0$
we get, up to three fermions terms:
\begin{eqnarray}
\delta \,\lambda^{{\mathcal I}1}&=&
 {\rm i}\,\nabla _ {\mu}\, z^{\mathcal I}
\gamma^{\mu} \epsilon^1 +\, W^{{\mathcal I}11}\epsilon _1
\label{chiral} \\
\delta \,\lambda^{{\mathcal I}2}&=&
 - G^{-{\mathcal I}}_{\mu \nu} \gamma^{\mu \nu}
\epsilon _1\,+\, g \, W^{{\mathcal I}21}\epsilon _1.
\label{gaugin}
\end{eqnarray}
From eqs. (\ref{chiral}) and (\ref{gaugin}) we immediately see
that the spinors $\l^{{\mathcal I} 1}$ transform into the scalars
$z^{\mathcal I}$ (and should therefore belong to $N=1$ chiral
multiplets) while the spinors $\l^{{\mathcal I} 2}$ transform into
the matter vectors field strengths $G^{-{\mathcal I}}_{\mu \nu}$
(and should then be identified with the gauginos of the $N=1$
vector multiplets). Let us decompose the world indices ${{\mathcal
I}}$ of the $N=2$ special-K\"ahler $\s$-model as follows:
${{\mathcal I}} \Rightarrow ( i , \a )$, with $ i =1,\cdots ,
n_C$, $\a =1,\cdots , n'_V=n_V-n_C$, where $n_C$ and $n'_V$ are
respectively the number of chiral and vector multiplets of the
reduced $N=1$ theory while $n_V$ is the number of $N=2$ vector
multiplets. By an appropriate choice of coordinates, we call $z^i$
the coordinates on $\cM_R $, $z^\a$ the coordinates on the
orthogonal complement. Then it is easy to see that the metric
$g_{{\mathcal I} \bar{\mathcal J}}$ has only components
$g_{i\bar\jmath}, g_{\a\bar\b}$, while $g_{i\bar\a}=0$. Then, if
we decompose the gauginos $\l^{{\mathcal I} 2} \Rightarrow (\l^{ i
2}, \l^{\a 2})$, the above truncation implies, by supersymmetry,
$\l^{ i 2}=0$ and, for consistency,
\begin{equation}\label{gau2}
\delta \,\lambda^{{ i}2}=
 - G^{-{ i}}_{\mu \nu} \gamma^{\mu \nu}
\epsilon _1\,+\, g \, W^{{ i}21}\epsilon _1 =0.
\end{equation}
Setting  $G^{-  i}_{\mu\nu}=0$ gives:
\begin{equation}\label{mattervectors}
  G^{-  i}_{\mu\nu} = -g^{ i  \bar{\mathcal J}} \nabla
  _{ \bar{\mathcal J}}\bar L^{{\bf\L}}  {\rm Im} \cN_{{\bf\L}\bf\S}
  F^{-\bf\S}_{\mu\nu}
=- g^{ i  {\bar\jmath}} \nabla
  _{ {\bar\jmath}}\bar L^{ \L}  {\rm Im} \cN_{ \L \S}
  F^{- \S}_{\mu\nu}=0
\end{equation}
implying
\begin{equation}
\nabla _{ {\bar\jmath}}\bar L^{ \L} =\bar f
  _{ {\bar\jmath}}^{ \L}=0.
  \label{dl0}
\end{equation}
Moreover, $W^{{ i}21}=0$ implies $ P^3_{\dot\L} =0$, $k^i_{\dot
\L} =0$. Note that the integrability condition of equation
(\ref{dl0}) is:
\begin{equation}\label{intdl0}
  \nabla_i \nabla_j L^{ \L} = {\rm i} C_{ij {\mathcal K}} g^{{\mathcal K}
  \bar{\mathcal K}}\nabla_{\bar{\mathcal K}} \bar L^{ \L} = {\rm i} C_{ij  k}
  g^{  k{\bar k}}\nabla_{ {\bar k}} \bar L^{ \L}+ {\rm i} C_{ij\a}
   g^{\a
  {\bar\a}}\nabla_{{\bar a}} \bar L^{ \L}=0.
\end{equation}
where $C_{ijk}$ is the 3-index symmetric tensor appearing in the
equations defining the special geometry (see {\it e.g.} ref.
\cite{cdf},\cite{abcdffm}). Since the first term on the r.h.s. of
equation  (\ref{intdl0}) is zero on $\cM_R $ (equation
(\ref{dl0})), equation  (\ref{intdl0}) is satisfied by imposing:
\begin{equation}\label{ccond1}
C_{ij\a}=0
\end{equation}
so that only the $N=2$ special-K\"ahler manifolds satisfying the
constraint (\ref{ccond1}) are suitable for reduction.
\par
Let us now consider the other condition coming from the reduction
of the gravitino transformation law, eq. (\ref{o21}). It implies,
on the curvature $\Omega_2^{\ 1}$ of the $SU(2)$ connection
$\omega_2^{\ 1}$:
\begin{equation}
\Omega_2^{\ 1} \,=\, -{\rm i}\,\IC_{\alpha\beta}{\mathcal
U}^{\alpha }_1 \wedge {\mathcal U}^{\beta 2} =0 \label{curform}
\end{equation}
Let us now note that the scalars in $N=1$ supergravity must lie in
chiral multiplets, and span in general a K\"ahler-Hodge manifold.
It is therefore required that the holonomy of the quaternionic
manifold be reduced: $
 {\rm Hol} \left({\cal M}^{Q}\right) \subset SU(2) \times Sp(2n_H) \to
 {\rm Hol} \left({\cal M}^{KH}\right) \subset U(1) \times
 SU(n)$.
Therefore the $SU(2)$ and the $Sp(2n_H)$ indices have to be
decomposed accordingly. We set $\a \to (I,\dot I )\in U(1) \times
SU(n_H) \subset Sp(2n_H)$ and the symplectic metric
$\IC_{\alpha\beta}$ reduces to  $\IC_{I\dot J} = - \,\IC_{\dot J
I}= \delta_{I\dot J}$.
 From equation  (\ref{curform}) we see that the constraint
(\ref{o21}) for involution is satisfied iff ${\mathcal
U}^{1I}\wedge {\mathcal U}^{1\dot I}=0 $ that is if, say, the
subset ${\mathcal U}^{2I}= \left({\mathcal U}^{1\dot I}\right)^*$
 of the quaternionic vielbein is set to zero on a
 submanifold $\cM^{KH}\subset \cM^Q$.
\\
When this condition is imposed, the submanifold $\cM^{KH}$ has
dimension at most half the dimension of the quaternionic manifold
and the $SU(2)$ connection is
 reduced to a $U(1)$ connection, whose curvature on $\cM^{KH}$ is:
\begin{equation}  \Omega^3|_{\cM^{KH}} = {\rm
i}\lambda{\mathcal U}^{1I}\wedge  {\mathcal U}_{1 I }= {\rm
i}\lambda{\mathcal U}^{1I}\wedge \overline{ {\mathcal U}^{1 I }}
\label{omega3}
\end{equation}
so that the $SU(2)$-bundle of the quaternionic manifold is reduced
to a $U(1)$-Hodge bundle for the $n_H$ dimensional complex
submanifold spanned by the $n_H$ complex vielbein ${\mathcal
U}^{1I}$. The truncation corresponds therefore to select a
$n_H$-complex dimensional submanifold $\cM^{KH} \subset \cM^{Q}$
spanned by the vielbein ${\mathcal U}^{1I}$ and to ask that, on
the submanifold, the $2n_H$ extra degrees of freedom are frozen,
that is: ${\mathcal U}^{2I}|_{\cM^{KH}} =\left({\mathcal U}_{1\dot
I}\right)^*|_{\cM^{KH}} \,=\,0 \label{zero}$.
\par
In order to consistently impose the constraint $\cU^{2I} =0$, it
is necessary to check its involution:
\begin{equation}
d{\mathcal U}^{2I} ={i\over 2}(\omega^1+{\rm i}\omega^2)  \wedge
{\mathcal U}^{1I}+{i\over 2}\omega^3 \wedge {\mathcal U}^{2I} -
\Delta^{I}_{\ J} \wedge {\mathcal U}^{2J} - \Delta^{I}_{\ \dot J}
\wedge {\mathcal U}^{2\dot J} =0.
 \label{tors2}
\end{equation}
Eq. (\ref{tors2}) implies that we must set to zero, for a
consistent reduction, also the ``off-diagonal'' part of the
$Sp(2n_H)$ connection, $\Delta^{I}_{\ \dot J}$. This in turn
implies a constraint on the four-fold symmetric tensor
$\Omega_{\a\b\gamma\delta}$ appearing  in the symplectic part of
the quaternionic curvature \cite{lungo}:
\begin{equation}
\Omega_{IJK\dot L}|_{\cM^{KH}} =0.
\end{equation}
\par
Finally, let us look at the reduction of the hypermultiplets
transformation laws, after imposing (\ref{o21}).
 They become,
after putting $\epsilon_2 =0$:
\begin{eqnarray}
&&{\mathcal U}_u^{1I}\delta\,q^u =  \bar {\zeta}^{I}
  \epsilon^1 \,; \quad
  {\mathcal U}_u^{2I}\delta \,q^u = -\IC^{I\dot J}\bar {\zeta}_{\dot J}
  \epsilon _1 \label{quat}\\
&&\label{iper1}
  \delta\,\zeta _{I}={\rm i}\,
{\mathcal U}^{2 \dot J}_{u}\, \IC_{I \dot J} \, \nabla _{\mu}\,q^u
\,\gamma^{\mu} \epsilon^1 \,+ g N^1_I \epsilon _1 =\left(\delta\,\zeta ^{I}\right)^*\\
&&\label{iper2} \delta\,\zeta _{\dot J}={\rm i}\, {\mathcal U}^{2
I}_{u}\, \IC_{\dot J I } \, \nabla _{\mu}\,q^u \,\gamma^{\mu}
\epsilon^1 \,+ g N^1_{\dot J} \epsilon _1 =\left(\delta\,\zeta
^{\dot J}\right)^*
\end{eqnarray}
Equations (\ref{quat}) show that the fermionic partners of the
retained scalars $\cU^{1I}$ are the spinors $\zeta_I$, while the
partners of the scalars $\cU^{2I}$ which have to be dropped out
are the spinors $\zeta_{\dot I}$. Consistency of the truncation
then imposes that the gauging contribution $g N^1_{\dot J}$ has to
be zero on the reduced $N=1$ theory.
\par
In the sequel, we just quote the main results on the gauged
theory, which are extensively discussed in \cite{lungo}.
 The truncation $N=2 \to N=1$
implies, on the gauged theory:
\\
- The $D$-term of the $N=1$-reduced gaugino $\lambda^\L = -2
f^\L_i \l^{i2} $ is:
\begin{equation}
D^\L = -2 f^\L_\a W^{\a 21} =-2g_{(\L)}({\rm Im}f)^{-1 \L\S}
\left(P^3_\S(w^s) +P^0_\S (z^i)\right)\label{d2}
\end{equation}
- The $N=1$-reduced superpotential, that is the gravitino mass, is
\begin{equation}
L(z,w)=\frac{\rm i}{2}g_{(X)}L^{X} \left( P^1_{X} - {\rm i}P^2_{X}
\right) \label{sp} \end{equation}
 and is  a holomorphic function of its
coordinates $z^i$ and $w^s$.\\
- The fermion shifts of the $N=1$ chiral spinors $\chi^i =\l^{i1}$
coming from the $N=2$  gaugini are $N^i \equiv g W^{i11} =2
g^{i\bar\jmath}\nabla_{\bar\jmath}\bar L$.\\
- The fermion shifts of the $N=1$ chiral spinors $\zeta^s=\sqrt{2}
P^{I,s} \zeta_I $ coming from $N=2$  hypermultiplets ($P^{I,s}$
are the scalar vielbein on $\cM^{KH}$) are $N^s =-4 g_{(X)}k^t_{X}
\bar L^{X} {\mathcal U}^{1\dot I}_t {\mathcal U}_{2\dot I}^s = 2
g^{s\bar s}\nabla_{\bar s}\bar L$.
 Furthermore, the consistency of the truncation of the second
gravitino multiplet $\delta \psi_{\mu 2} =0$ and of the spinors
$\zeta_{\dot I}$ in the hypermultiplets sector for the gauged
theory gives the gauging constraints:
\begin{eqnarray}
\hat \omega_1^{\phantom{1} 2}=0 &\Longrightarrow & g_{({\bf \L })}
A^{{\bf \L }} \left(P^1_{{\bf \L }} -{\rm i} P^2_{{\bf \L
}}\right)= 0  \Rightarrow g_{({\L })} A^{{\L }} \left(P^1_{{\L }}
-{\rm i} P^2_{{\L }}\right)= 0
\label{orto1}\\
S_{12}=0 &\Longrightarrow & g_{({\bf \L })} L^{{\bf \L }}
P^3_{{\bf \L }} =0 \Rightarrow g_{({X })} L^{{X }} P^3_{{X }} =0
\label{orto2}\\
\delta \zeta_{\dot I} = 0 & \Longrightarrow & g_{({\bf \L } )}
k^s_{{\bf \L }} \bar L^{{\bf \L }} = 0 \Rightarrow g_{(X)} k^s_{X}
\bar L^{X} = 0 \label{orto3}
\end{eqnarray}
If we call $G^{(2)}$ the gauge group of the $N=2$ theory and
$G^{(1)}\subseteq G^{(2)}$ the gauge group of the corresponding
$N=1$ theory, then we have that the adjoint representation of
$G^{(2)}$ decomposes as $Adj(G^{(2)}) \Rightarrow  Adj(G^{(1)}) +
R(G^{(1)})$, where $R(G^{(1)})$ denotes some representation of
$(G^{(1)})$. The gauged vectors of the $N=1$ theory are restricted
to the subset $\{A^{ \L}\}$ generating $Adj(G^{(1)})$ (that is,
with the same decomposition of the ungauged theory, we set ${\bf
\L } \to (\L , X )$, with $\L\in Adj(G^{(2)}$ and $X \in
R(G^{(1)})$).
 The quaternionic Killing vectors of the $N=2$ theory then decompose
as $k^u_{{\bf \L }} \Rightarrow \{k^s_{\L}, k^{\bar s}_{\L},
k^t_{\L}, k^s_{X}, k^{\bar s}_{X}, k^t_{X}\}$.
 Obviously, we must have that $k^s_{X} =0$ since  the
Killing vectors of the reduced submanifold have to span the
adjoint representation of $G^{(1)}$. Viceversa, the Killing
vectors with world index in the orthogonal complement, $k^t_{{\bf
\L }}$, must obey $k^t_{\L} =0$, while $k^t_{X}$ are in general
different from zero. Indeed, the isometries generated by
$k^t_{{\bf \L }}$ would not leave invariant the hypersurface
describing the submanifold ${\cal M}^{KH} \subset {\cal M}^Q$.
Since we have found that $k^s_{X}=0$, equation (\ref{orto3}) is
identically satisfied. Eq.s (\ref{orto1}) and (\ref{orto2}) are
satisfied by requiring: $ P^1_\L = P^2_\L =0 \, ; \quad
P^3_{X}=0$.
\par
 We are left with an $N=1$ theory
coupled to $n'_V$ vector multiplets ($ \L =1,\cdots ,n'_V$) and
$n_C +n_H$ chiral multiplets ($X =0,1,\cdots ,n_C$) with
superpotential (\ref{sp}). All the isometries of the scalar
manifolds are in principle gauged since the D-term of the reduced
$N=1$ theory depends on $P^0_\L (z,\bar z )+ P^3_\L (w,\bar w )$.

%%%%%%%%%%%%%%%%%%%%%%
%%%%%%%%%%%%%%%%%%%%%%

\end{document}